\documentclass{elsart}
\usepackage[english]{babel} 
\usepackage{times}
\usepackage{graphicx}
\usepackage{paralist}
\usepackage{cite} 
\begin{document} 
\begin{frontmatter}

\title{The role of water in the behavior of wood} 

\author[a1]{Dominique Derome\corauthref{cor}},
\corauth[cor]{Corresponding author. \"Uberlandstrasse 129, CH-8600 D\"ubendorf}
 \ead{dominique.derome@empa.ch}
\author[a1,a2]{Ahmad Rafsanjani}
\author[a3]{Stefan Hering}
\author[a1,a2]{Martin Dressler}
\author[a1,a2]{Alessandra Patera}
\author[a4]{Christian Lanvermann}
\author[a1]{Marjan Sedighi-Gilani}
\author[a3]{Falk K. Wittel}
\author[a4]{Peter Niemz} and
\author[a1,a5]{Jan Carmeliet}
\address[a1]{Laboratory for Building Science and Technology, Swiss Federal Laboratories for Materials Science and Technology, EMPA, D\"ubendorf, Switzerland}
\address[a2]{Department of Civil, Environmental and Geomatic Engineering, ETH Zurich, Zurich, Switzerland} 
\address[a3]{Institute for Building Materials, ETH Zurich, Zurich, Switzerland} 
\address[a4]{Wood Physics Group, Institute for Building Materials, ETH Zurich, Zurich, Switzerland} 
\address[a5]{Chair of Building Physics, ETH Zurich, Zurich, Switzerland} 

\begin{keyword}
Wood, coupling, water, mechanical behavior, poromechanics, multiscale, imaging methods
\end{keyword}
 \begin{abstract}
Wood, due to its biological origin, has the capacity to interact with water. Sorption/desorption of moisture is accompanied with swelling/shrinkage and softening/hardening of its stiffness. The correct prediction of the behavior of wood components undergoing environmental loading requires that the moisture behavior and mechanical behavior of wood are considered in a coupled manner. We propose a comprehensive framework using a fully coupled poromechanical approach, where its multiscale implementation provides the capacity to take into account, directly, the exact geometry of the wood cellular structure, using computational homogenization. A hierarchical model is used to take into account the subcellular composite-like organization of the material. Such advanced modeling requires high resolution experimental data for the appropriate determination of inputs and for its validation. High-resolution x-ray tomography, digital image correlation, and neutron imaging are presented as valuable methods to provide the required information.
\end{abstract}
\end{frontmatter}

%%%%%%%%%%
\section{Introduction}
Wood, an orthotropic cellular material, has the thought-provoking property of adsorbing water molecules from its surrounding into its hierarchical material structure. Although known and employed for millenniums, wood in use and exposed to environmental loading is subject to varying moisture contents (MCs) and consequently exposed to a rich pallet of moisture-induced processes, possibly leading to wood degradation.

Wood, compared to other materials, is very hygroscopic. As water molecules attach themselves to the hydrophilic matrix in the cell walls, the induced fluid- solid-gas interaction forces result in a swelling of the cell walls. Due to cell geometrical irregularities and latewood and earlywood swelling incompatibilities, moisture-induced internal stresses originate and highly influence the hygromechanical behavior of wood as observed at the macroscale. Wood is known as a material showing a high influence of mechanical forces on water sorption. The interaction of the moisture and mechanical behavior of wood is best observed in swelling. Absorption of moisture in wood, in the hygroscopic range, that is, until around 30\% MC mass per mass, results in swelling and reduced stiffness. The microscopic origin of this behavior lies at the scale of the cell wall. The cell wall material is composed in almost almost equal quantity of stiff cellulose microfibrils and a soft polymeric matrix. The hydrophobic crystalline cellulose is surrounded by hydrophilic amorphous cellulose, immersed in a hydrophilic amorphous matrix, of hemicelluloses bound by lignin. The sorption of water molecules in between the hydrophilic molecules pushes the constituents apart, resulting in swelling and a reduction of stiffness of the matrix. The thin internal and external cell wall layers (i.e. S$_3$ and S$_1$) act as corsets due to the winding of the cellulose fibrils around the cell. In the central and, by far, thickest cell wall layer, namely S$_2$, the cellulose microfibrils are almost parallel to the longitudinal axis of the cells, although the presence of an angle (called 'microfibril angle' (MFA)) results in a helicoidal organization of the fibrils. Although the effect of varying the MFA from 0$ \circ$ to 30$^\circ$ on the transverse stiffness and swelling properties of softwoods is small, during the sorption of moisture, the general orientation of the microfibrils in the S$_2$ layer results in notable swelling in the transverse directions of the cell and almost none along the longitudinal direction. Furthermore, the helicoidal organization of the cellulose fibrils in the cell wall causes swelling to be more pronounced normal to the cell wall than along the cell wall direction.
\begin{figure}[htb] \centering{ \includegraphics[width=8.cm]{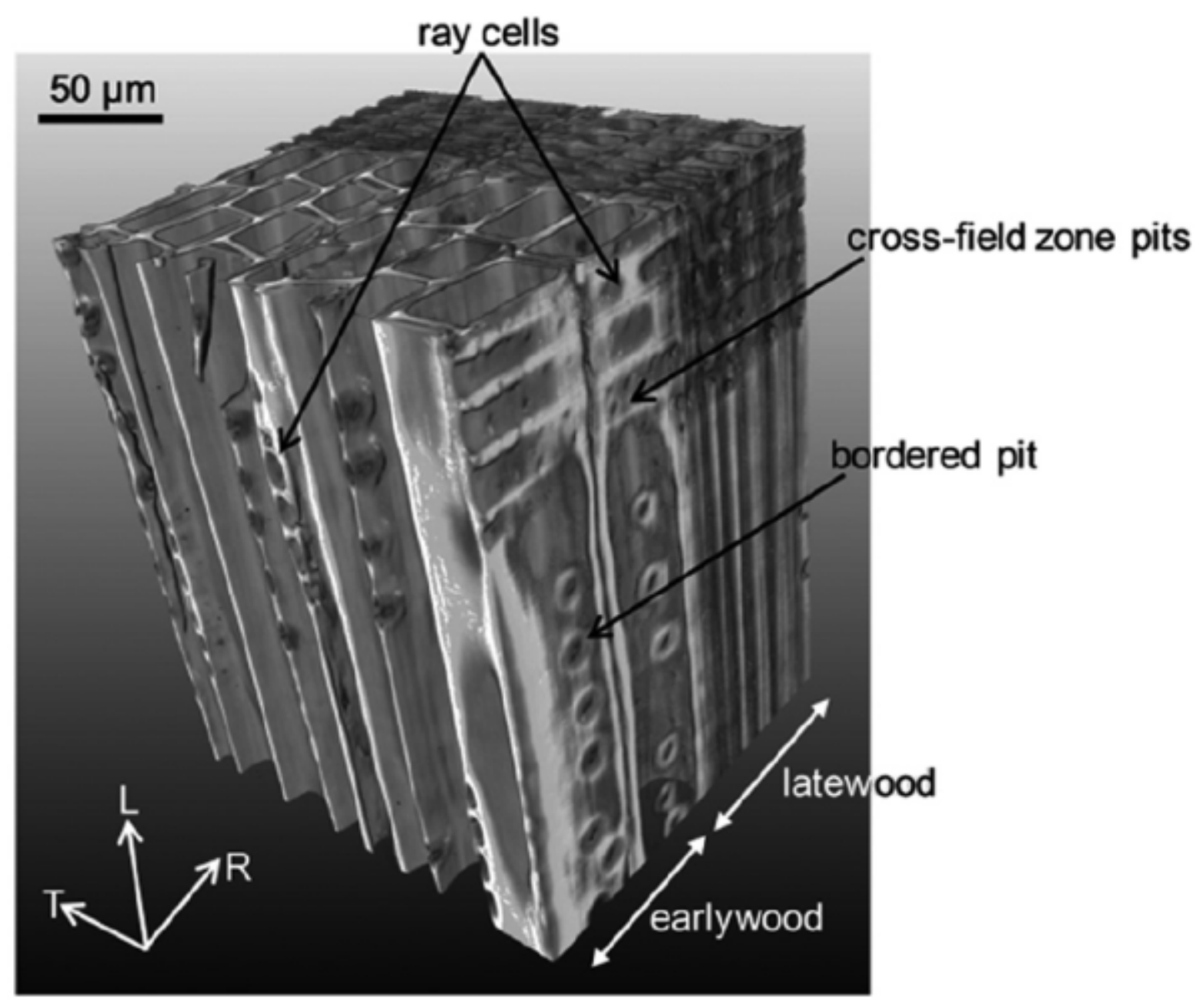} }
 \caption{Tomographic image of spruce highlighting different anatomical features of softwood.} 
\end{figure}

Although swelling originates from the cell wall level, its anisotropic nature seems to find its origin mainly at the cellular architecture level (see Figure 1). On a mesoscopic scale, wood consists mostly of longitudinal tracheid cells and also of radially oriented ray cells, the latter representing 5\% of the wood volume. Across the growth ring, the thin-walled earlywood cells with large internal cavities, called 'lumens', often gradually change to thick-walled latewood cells with small-sized lumens. The variation of the cellular structure across the growth ring has been found responsible for the anisotropy of swelling and stiffness (Boutelje, 1962; Watanabe et al., 2000). Furthermore, recent findings on the isotropic swelling of latewood and anisotropic swelling of earlywood in the transverse plane indicate the possibility of latewood/earlywood interaction (Derome et al., 2011).

At present, we do not completely understand the origins and play of water in the moisture, mechanical, and time-dependent processes in wood, not to mention their hysteresis (e.g. as investigated for paper by Derluyn et al., 2007) or its impact of durability (e.g. Viitanen et al., 2010). We present our current work in the understanding of the role of water in the behavior of wood. A combined modeling/ experimental approach is used to study further the role of water in these processes. A multiscale approach, using modeling and advanced experimental techniques, looks at the moisture-related properties of wood at different scales. The poroelastic modeling is used and upscaling procedures are developed from cell wall constituents to the scales of cell wall, cell, latewood and earlywood, growth ring, and wood timber. To document and validate the modeling efforts, a comprehensive experimental plan has been established where synchroton X-ray tomographic datasets document the complex swelling/shrinkage behavior of wood cells, swelling at the growth ring scale is documented with digital image correlation (DIC), and finally, neutron projections document changes in MC at the growth ring level.
%%%%%%%%%%%
\section{Modeling the water and mechanical behavior of wood}
We propose a comprehensive framework using a fully coupled poromechanical approach, where its multiscale implementation provides the capacity to take into account, directly, the exact geometry of the wood cellular structure, using computational homogenization. An hierarchical model is used to take into account the subcellular composite-like organization of the material.
%%%
\subsection{Poromechanical approach}
A rigorous way of taking into account the interaction of fluids with the solid matrix in porous materials is based on the theory of poromechanics, introduced by Biot (1941). Within the context of thermodynamics of open porous continua, Coussy (2010) presented a general framework to formulate adequate constitutive equations for poroelastic behavior. The poromechanical approach starts from an energy viewpoint where stress/strain and chemical potential/MC are seen as pairs determining the energy content. With one equation expressing the energy content of the whole system, any change in stress field or chemical potential leads to a coupled response in both the strain and the MC. Thus, the poromechanical approach is suited to cover coupled effects between the mechanical and moisture response as observed in wood. A full description of the model described below is found in Carmeliet et al. (2013).
\begin{figure}[htb] \centering{ \includegraphics[width=12.cm]{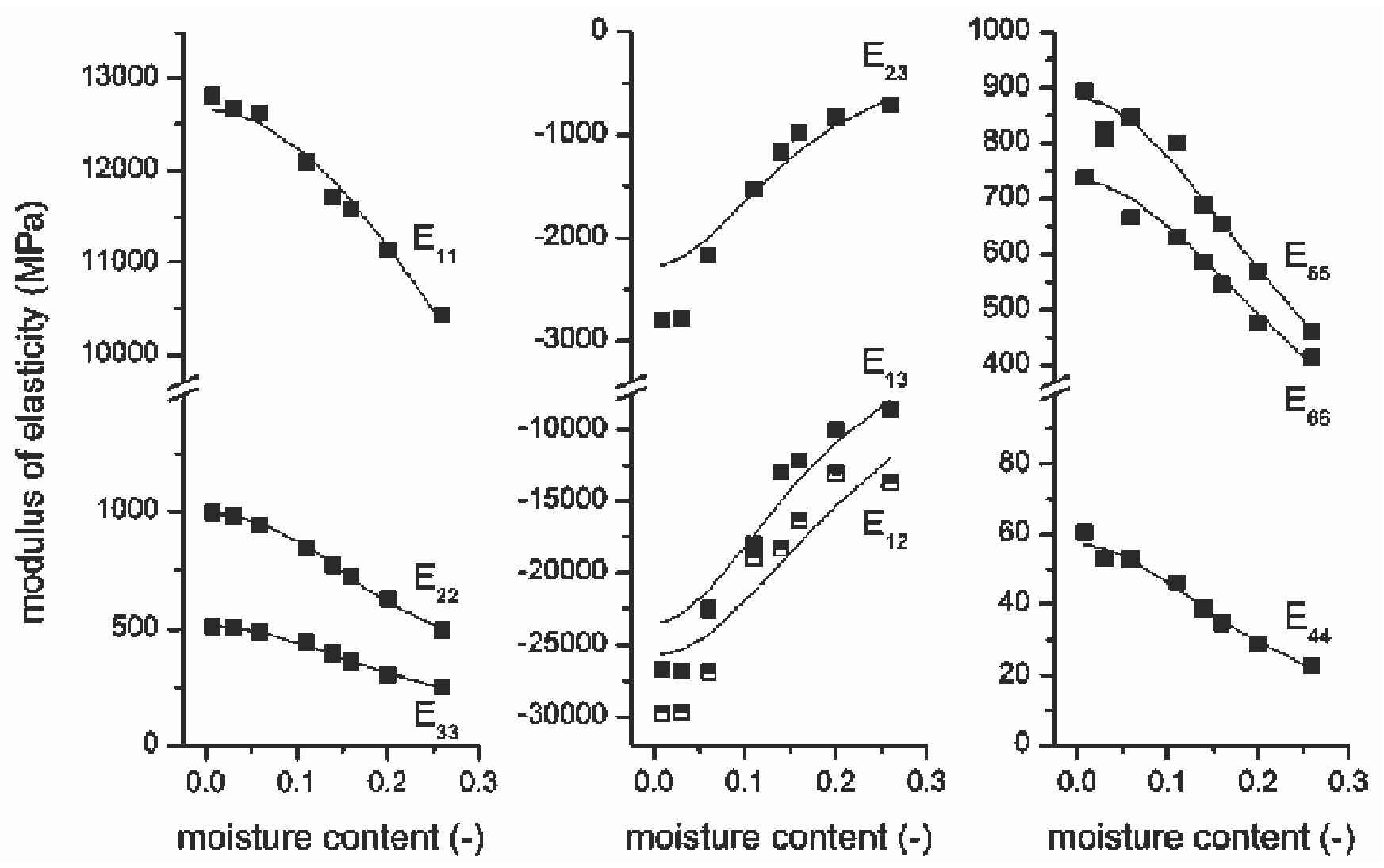} }
 \caption{Poromechanical model results (continuous lines) fitted with experimental modulus of elasticity (black square) reported by Neuhaus (1981).} 
\end{figure}

In this coupled poromechanical approach, the free energy content, $\Phi$ (J/m$^3$), of a porous solid is described by the contributions of strain and moisture and the interaction between strain and moisture. We can describe this system in terms of the applied fields, that is, stress and chemical potential, as follows
\begin{equation}
	d\Phi=\varepsilon_{ij}d\sigma_{ij}+nd\mu
\end{equation}
where $\varepsilon_{ij}$ (m/m) and $\sigma_{ij}$ (N/m$^2$) denote the strain and stress tensors, respectively; $n$ (mol/m$^3$) and $m$ (J/mol) are the mole number density of the material and chemical potential of the adsorbed moisture, respectively.

The partial differentials of equation (1) yield two constitutive equations that describe the coupled mechanical and moisture behavior of wood. The first of these two constitutive equations (equation (2)) describes the strain ($\varepsilon_{ij}$) with respect to the stress ($\sigma_{ij}$) and the chemical potential ($\mu$). The second constitutive equation (equation (3)) gives the MC $u$ with respect to stress and chemical potential as follows 
\begin{eqnarray}
	d\varepsilon_i &= C_{ij}d\sigma_j+B_id\mu\\
	du &=\frac{\tilde{\rho}}{\rho}(B_id\sigma_i+Md\mu)
\end{eqnarray}
where $C_{ij}$ is the compliance (m$^3$/Nm), $B_i$ is the coupling coefficient (mol/J), $M$ is the moisture capacity(mol$^2$/m$^3$/J), while $\rho$ (kg/m$^3$) and $\tilde{\rho}$ (kg/mol) stand for volume density of the dried wood and molar mass of water, respectively.

Literature data were used to determine the introduced parameters. Table 1 lists the fitted parameter. Figure 2 shows the capacity of the model to capture the full range of coupled moisture/mechanical behavior of wood.
\begin{table*}[htb]\label{tab1}
\caption{Parameters fitted to the model using data from Neuhaus (1981) for spruce. The scaling value $C_{11}^0$ = 1/12655 MPa is the compliance in the longitudinal direction at zero moisture content}
  \begin{tabular}{p{3cm}p{3cm}p{3cm}p{3cm}} \hline
    $ij$& $\alpha_{ij}=\frac{C_{ij}^{[0]}}{C_{11}^{[0]}}$& $\beta_{ij}=\frac{B_{ij}^{[1]}}{C_{ij}^{[0]}}$& $\_i^{[0]}$\\ \hline
    11 (LL) & 1 & 3 & 0.33 \\
    22 (RR) & 13 & 15 & 0.14 \\
	  33 (TT) & 25 & 16 & 0.31 \\
		12 (LR) & -0.5 & 17 & - \\
		13 (LT) & -0.5 & 29 & - \\
		23 (RT) & -5.6 & 38 & - \\
		44 (RT) & 222 & 24 & 0 \\
		55 (LT) & 14 & 14 & 0 \\
		66 (LR) & 17 & 12 & 0 \\\hline
  \end{tabular}
\end{table*}
%%%
\subsection{Multiscale approach}
The interaction of the moisture and mechanical fields within a multiscale modeling framework requires a comprehensive treatment. The appropriate coupling between the scales is not straightforward. A rigorous multiscale framework in the context of a homogenization theory for the hygromechanical analysis of hierarchical cellular materials is thus sought. Given that the growth ring behavior is easily accessible and is considered here as the macroscale for wood, the effort is to develop an upscaling technique from the cellular structure to the growth ring level. A detailed description of this work is found in Rafsanjani et al. (2012). 
\begin{figure}[htb] \centering{ \includegraphics[width=12.cm]{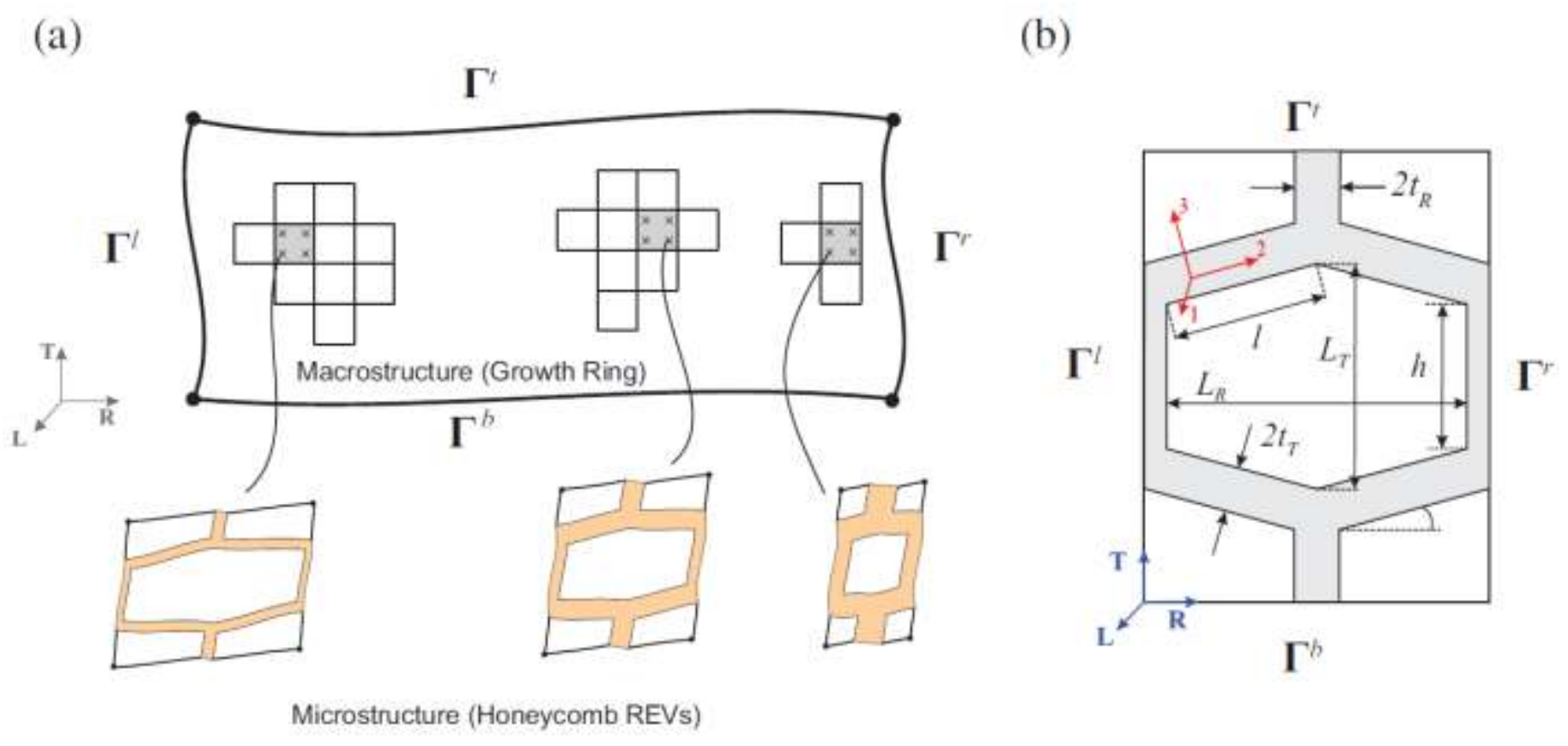} }
 \caption{(a) Two-scale upscaling scheme and underlying REVs corresponding to different elements and (b) geometry of REVs. (REVs: representative elementary volumes.)} 
\end{figure}
We consider a problem of infinitesimally small deformations with moisture induced swelling for growth rings in softwood as a cellular porous solid within the framework of the computational homogenization. Since the longitudinal dimensions of the wood cells are very long in comparison to the radial and tangential directions, the problem can be reduced to the analysis of the cross-section of the material, which is in a state of generalized plane strain. Verifying for separation of scales, the characteristic size of wood cells (microscale) is much smaller ($\approx$40 $\mu$m) than the growth ring (macroscal ) length scale ($\approx$3 mm). In the regions far away from the center of the stem of the tree, growth rings are periodically arranged in radial direction, which justifies an assumption on global periodicity of the timber. At the microscale, we assume local periodicity, where although wood cells have different morphologies corresponding to different relative ring positions, each cell repeats itself in its vicinity. Honeycomb representative elementary volumes (REVs) are selected for this current analysis. The concept of local and global periodicities in softwood, the proposed two-scale upscaling scheme, and the geometry of the REVs are illustrated in Figure 3.

\begin{figure}[htb] \centering{ \includegraphics[width=12.cm]{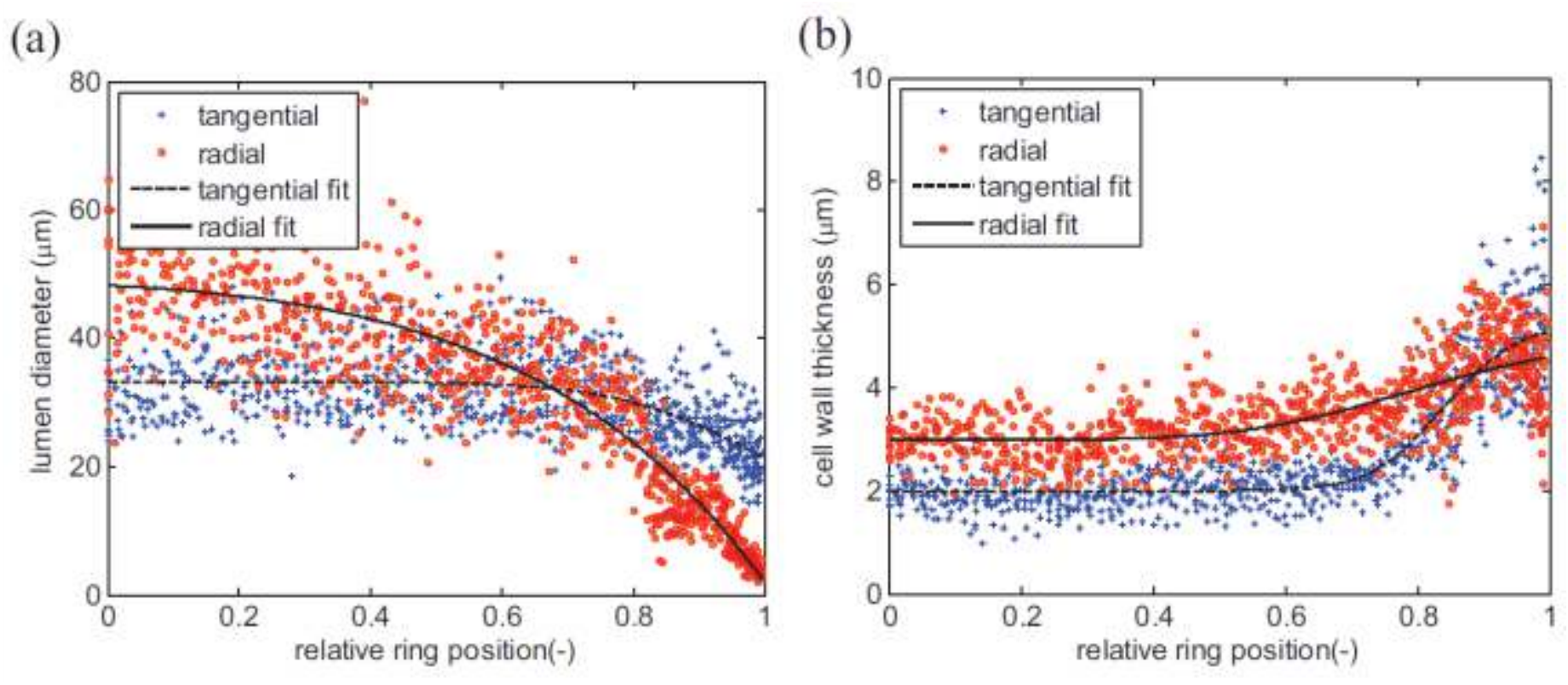} }
 \caption{(a) Lumen diameter and (b) cell wall thickness in radial and tangential directions as a function of the relative ring position, from SEM measurement on a spruce growth ring. (SEM: scanning electron microscopy.)} 
\end{figure}
We investigate the distribution of transverse anisotropy in swelling and mechanical behavior of wood from the cellular scale to the growth ring level by means of a finite element-based multiscale approach. The mechanical fields at the macroscopic level (growth ring) are resolved through the incorporation of the microstructural (cellular structure) response by the computational homogenization of different REVs of wood cells (honeycomb REVs) selected from a morphological analysis of wood at the cellular scale (Derome et al., 2012). Example of these measurements is shown in Figure 4.

These measurements are fitted to functions, which are used to predict the mechanical properties of unit cells along the growth ring using a master node homogenization technique. Figures 5 and 6 compare the calculated results to experimental values. Further comparison of the results of the predicted swelling and elastic moduli at the growth ring level with experimental data, although not reported here, is also very satisfactory (Rafsanjani et al., 2012).
\begin{figure}[htb] \centering{ \includegraphics[width=8.cm]{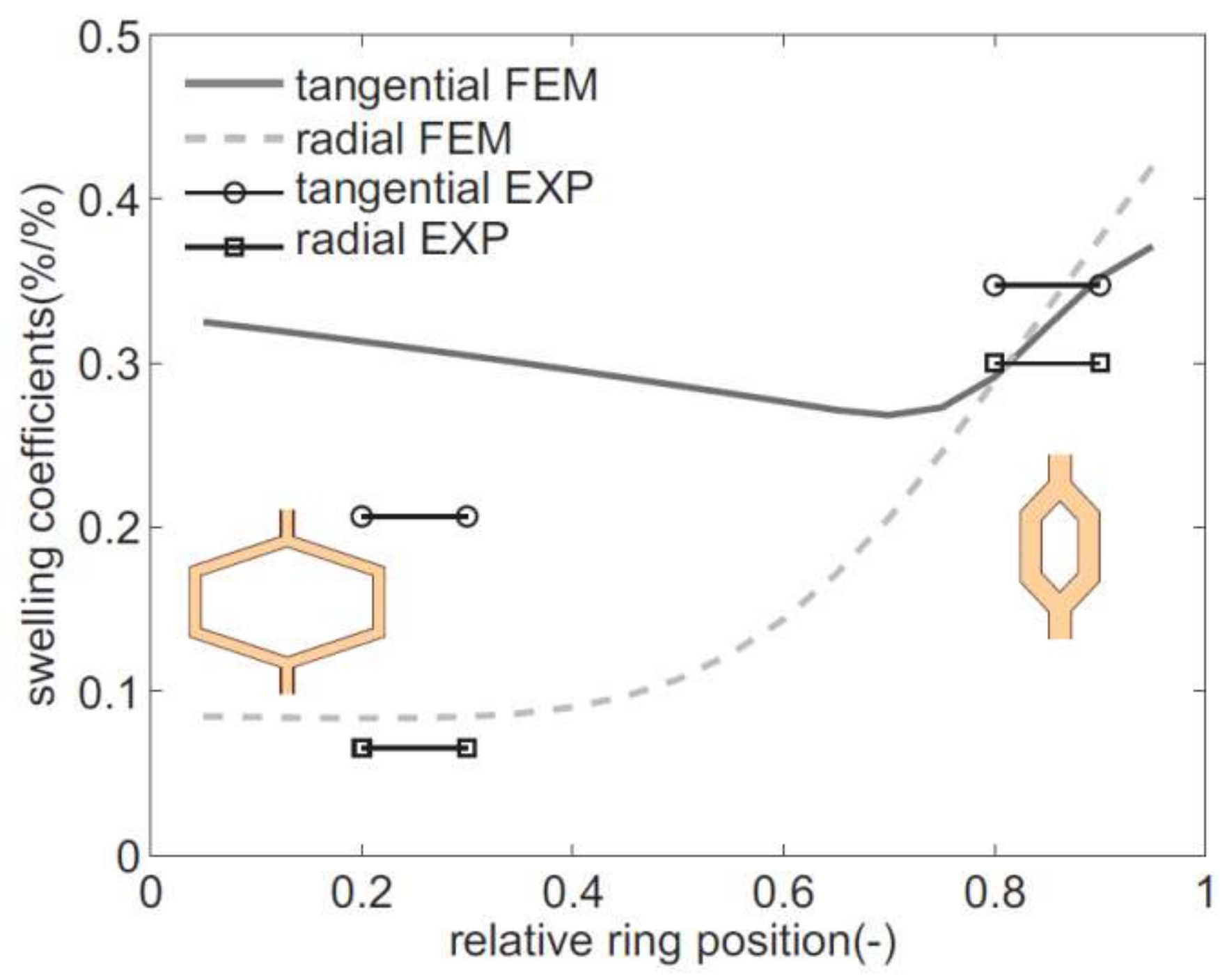} }
 \caption{Homogenized tangential and radial swelling coefficients for honeycomb unit cells (FEM finite element modeling, EXP experimental data from Derome et al. 2011).} 
\end{figure}
\begin{figure}[htb] \centering{ \includegraphics[width=8.cm]{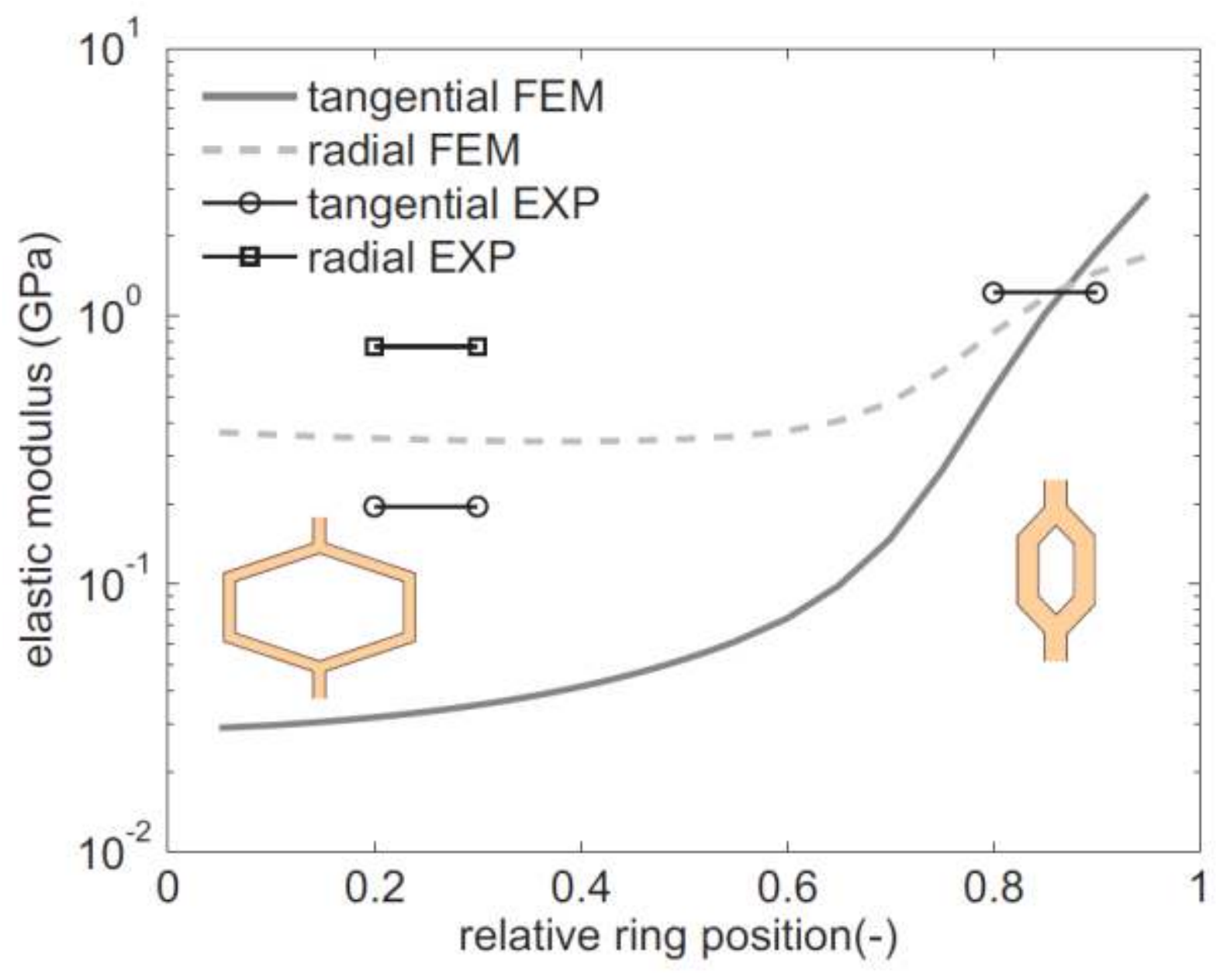} }
 \caption{Homogenized tangential and radial elastic moduli for honeycomb unit cells (FEM finite element modeling, EXP experimental data from Faruggia and Perre, 2000).} 
\end{figure}

This multiscale approach achieves a more realistic characterization of the anisotropic swelling and mechanical behavior of wood and, in particular, at describing the latewood and earlywood interaction caused by the material heterogeneity at the cellular scale.
%%%
\subsection{Hierarchic micromechanical approach}
As mentioned above, wood interaction with moisture, although influence by all scales, stems from the moisture sorptive capacity of its constituents, that is, hemicelluloses, celluloses, and lignin. Therefore, cell wall mechanical and moisture properties are not constant but depend on MC and the local spatial organization of the constituents. Even though the knowledge on the mechanical and moisture properties of constituents is rather sparse, modeling can be used to validate fundamental assumptions on constituents and their hierarchic organization. In a nutshell, in the wood cell wall, crystalline celluloses are surrounded by hemicelluloses, which are embedded in lignin, the latter binding together the cellulose/hemicellulose microfibrils (Figure 7(a)). 
\begin{figure}[htb] \centering{ \includegraphics[width=12.cm]{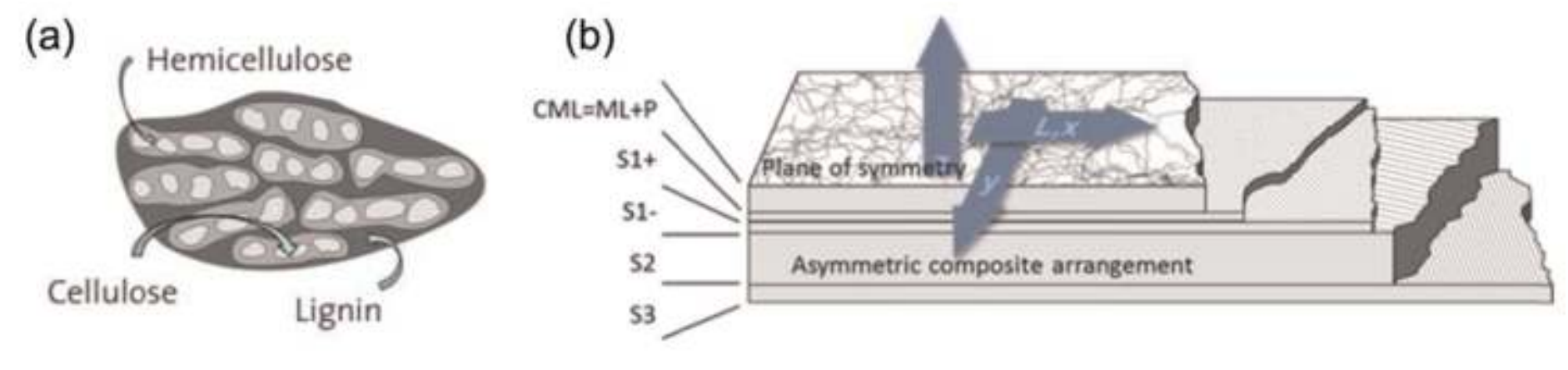} }
 \caption{(a) Constituents of a cell wall material and (b) cell wall of tracheid cells. ML: middle lamella; P: primary cell wall; CML: compound middle lamella.} 
\end{figure}
Microfibrils can be oriented unidirectionally or randomly in different cell wall layers. The angle between the longitudinal axis of a tracheid and the microfibrils is called microfibril angle (MFA). Five layers, namely, the middle lamella (ML), primary cell wall (P), and the secondary ones (S$_1$, S$_2$, and S$_3$), build up the cell wall of the tracheid (Figure 7(b)). We focus on the hygromechanical behavior of cell walls.
\begin{figure}[htb] \centering{ \includegraphics[width=12.cm]{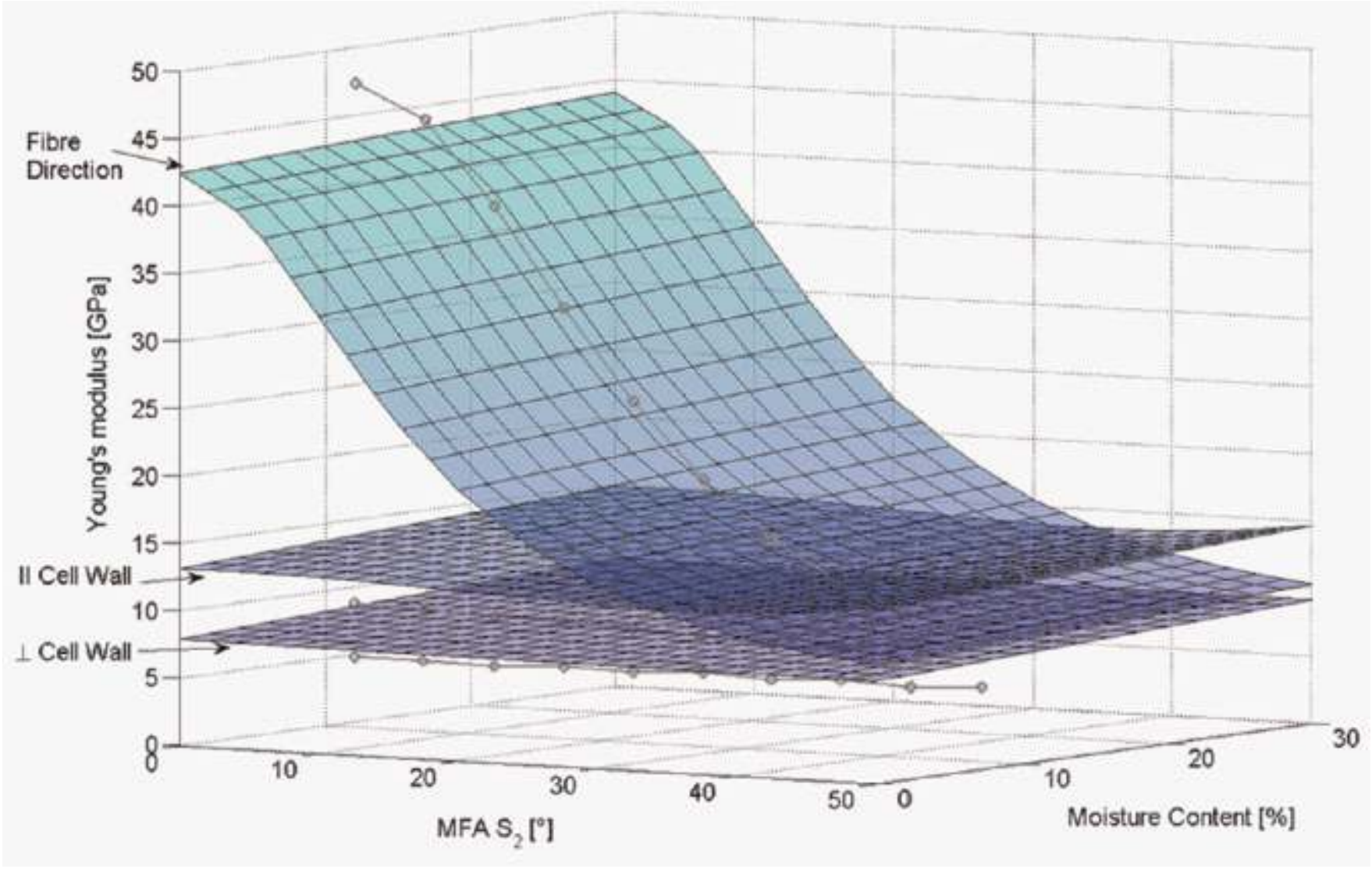} }
 \caption{Calculated moisture- and microfibril-dependent elasticity of the cell wall compared with the results proposed by Persson (2000) at a moisture content of 12\% (gray lines with circle markers).
MFA: microfibril angle.} 
\end{figure}
\begin{figure}[htb] \centering{ \includegraphics[width=12.cm]{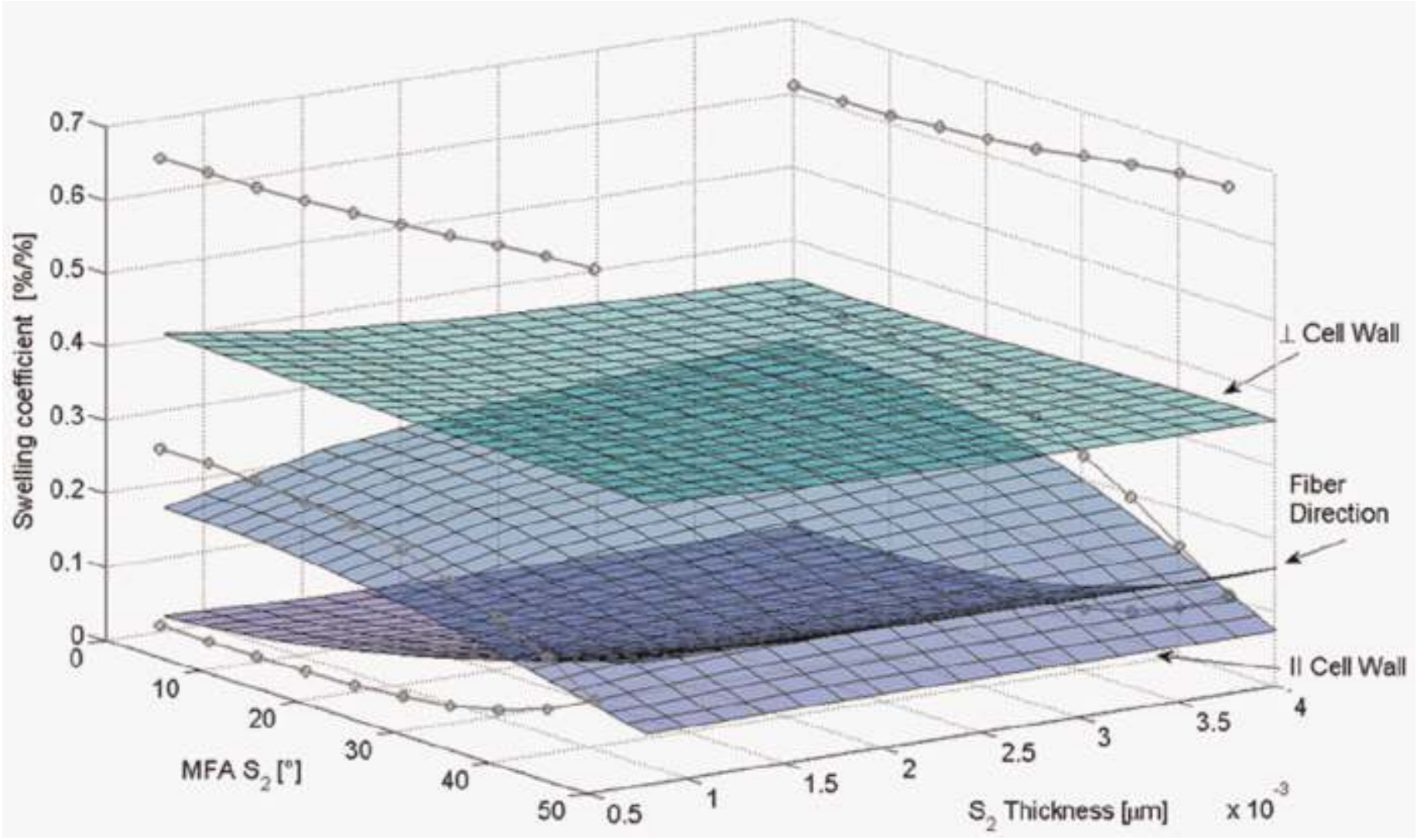} }
 \caption{Calculated swelling/shrinkage coefficients of the cell wall depending on the MFA and the thickness of the S$_2$ layer compared with the results proposed by Persson (2000) (gray lines with circle markers). MFA: microfibril angle.} 
\end{figure}

The layered nature of the cell walls, as well as the high aspect ratio of microfibrils, calls for a modeling approach analogous to the one used to model fiberreinforced composites. The first of the two main ingredients is the mixing rule, meaning the derivation of effective layer properties from the constituent properties and, as such, a homogenization step. For this purpose, we employ a modified Halpin Tsai approach with hygroexpansion. The second ingredient is the reduction of the five orthotropic layers to an equivalent single layer using classical laminate theory with the elastic properties of the constituents and the coefficients of hygroexpansion taken from Cave (1978) and Persson (2000). While our laminate approach follows Tsai and Hahn (1980), the mixture rules need to be discussed in detail. 

In comparative studies of mixture rules (Chamis, 1984), the Halpin-Tsai equations show the best predictive behavior. This is due to the fact that properties are not derived from first principles but are calculated using an empirical geometry factor for the spatial organization. This factor is estimated by fits to experimental data or detailed numerical models. For unidirectional (i.e. layers S$_1^{+/-}$, S$_2$, and S$_3$) and randomly oriented (i.e. layers ML and P) materials, these factors are known (Halpin and Kardos, 1976). One peculiarity of microfibrils is that the homogenization needs to be done in two steps, by combining first celluloses and hemicelluloses and then embedding this combination in lignin (Figure 7(a)). The two consecutive homogenization steps for the coefficients of hygroexpansion ( $\alpha$) were performed for the transversal isotropic $S_i$ layers as follows
\begin{equation}
	\alpha_l^i=\frac{V_f^i\alpha_f^iE_f^i+V_m^i\alpha_m^iE_m^i}{V_f^iE_f^i+V_m^iE_m^i}
\end{equation}
and for the longitudinal cell direction and for the directions perpendicular to the cells as follows
\begin{equation}
	\alpha_t^i=\alpha_f^i\sqrt{(V_f^i)}+(1-\sqrt{(V_f^i)})\left(\frac{1+V_f^i\nu_m^iE_f^i}{V_f^iE_f^i+V_m^iE_m^i}\right)\alpha_m^i
\end{equation}
where subscripts $f$ and $m$ denote the fiber phase and matrix phase, respectively, and $V$ is the corresponding volume fraction. Due to their random-oriented character, the ML and P layers were subsequently averaged considering Young's moduli in plane ($E_1,~E_2$) given by
\begin{equation}
	\alpha_{ro}=\frac{1}{36}\cdot \sum_{r=1}^{36}\frac{\alpha_lE_1\cos(\frac{r\pi}{18})+\alpha_tE_2\cos(\frac{r\pi}{18})}{E_1+E_2}
\end{equation}
Our moisture-dependent analytical model is validated against a computationally expensive numeric two-scale finite element model proposed by Persson (2000) for one specific MC as a function of the MFA. Note that we used composite sections of ABAQUS for this purpose for simplicity and that the hygroexpansion can be treated similarly to thermal expansion. Figure 8 provides the calculated elastic properties of the cell wall, which are despite of small deviations in good agreement with Persson (2000) in all material directions.

To estimate the free swelling of a piece of cell wall, a block of two adjacent cell walls is calculated. The swelling coefficients in the main directions of the cell wall are given in Figure 9. It is not surprising that the swelling in thickness direction dominates the one seen in the other orientations. Moreover, the dependency on MFA is more pronounced in the longitudinal direction and in the direction parallel to the cell wall. Compared to Persson (2000), all tendencies can be confirmed, even though the absolute values perpendicular to the cell wall differ. The latter can be explained by different mixing rules used at the microfibril level. For consistency, we employ Halpin-Tsai as described above, while Persson (2000) uses a numeric homogenization scheme from a so-called base cell with adjusted geometry.

In addition to provide the cell wall properties, this approach could be further used to study early/latewood for different cell wall and geometry properties in order to identify the factors dominating the swelling anisotropy of wood seen at the macroscale.
%%%%%%%%%%%%%%%%%%%%%
\section{Imaging methods}
The models presented above require high-resolution experimental data for the appropriate determination of inputs and for their validation. High-resolution tomography, DIC, and neutron imaging are presented as valuable methods to provide the required information.
\begin{figure}[htb] \centering{ \includegraphics[width=8.cm]{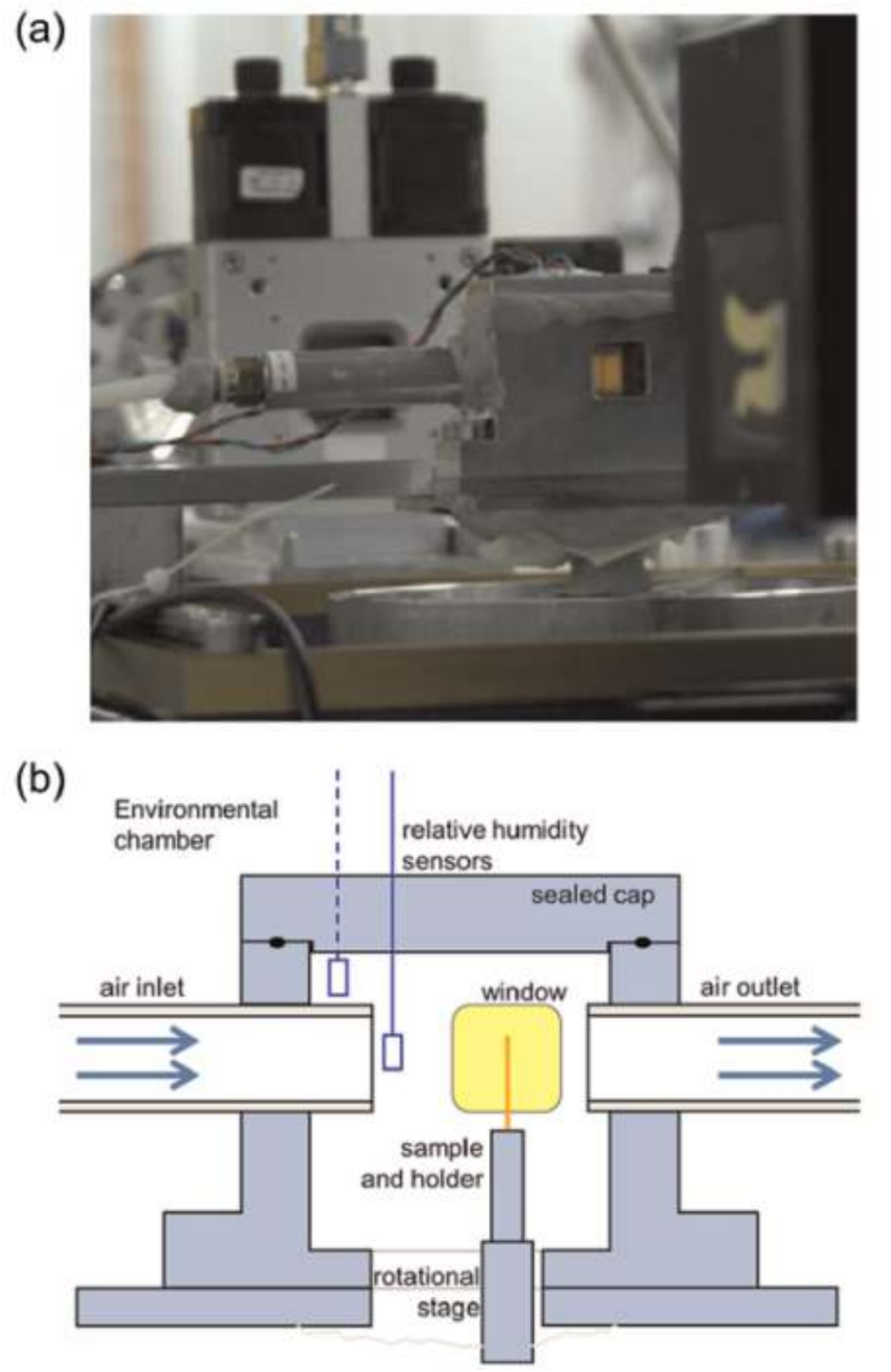} }
 \caption{(a) View of the experimental setup in the beamline and (b) schematic view of the environmental chamber.} 
\end{figure}
%%%%
\subsection{Swelling of wood probed by phase-contrast X-ray synchroton tomography}
We investigated the three-dimensional (3D), microscopic, dimensional changes of spruce (Picea abies (L. Karst)) wood samples due to controlled steps of the ambient relative humidity (RH; see Figure 10). This study was performed at the wood cellular scale by high-resolution synchroton radiation phase-contrast X-ray tomographic microscopy (srPCXTM), at the Tomcat beamline of the SLS of the Paul Scherrer Institute in Switzerland.

Due to the low X-ray attenuation coefficient of wood, the phase-contrast method is used in order to reconstruct the information related to the X-ray index of refraction. In fact, the difference in density at the boundary between air and wood is a source of phase contrast. The reconstruction uses information from the coherent X-ray propagation through the specimen and interference of the X-ray photons coming out of the specimen. Phase contrast allows a more accurate determination of the boundaries between the components.

Tomographic images of 500 $\mu$m x 500 $\mu$m x 8 $\mu$m samples, one of pure latewood and the second of pure earlywood, were taken as seen in Figure 11. Samples were let to achieve moisture equilibrium at five adsorption (starting at 25\%RH to 85\%RH) and four desorption (down to 10\%RH) steps plus a final adsorption step to 25\%RH. Affine registration of the dataset led to the identification of the orthotropic swelling strains, as shown in Figure 12, where the closure of the hysteretic loops demonstrates the reversibility of the process, and the start of a secondary loop, with the 25\%-10\%-25\%RH steps, provides further evidence of the hysteretic behavior. More details in the data processing that led to Figures 11 and 12 are found in Derome et al. (2011).

Figure 12 shows that, for spruce latewood, swelling and shrinkage are found to be larger, more hysteretic, and more homomorphic than for earlywood. Furthermore, while latewood undergoes similar strains in the transverse directions, the earlywood radial strains are less than a third of the tangential strains. The less homomorphic and smaller swelling/shrinkage of earlywood in radial direction is found to be caused by the presence of rays. The investigation of swelling/shrinkage strains was combined with dynamic vapor sorption (DVS) analysis, in which the samples were subjected exactly to the same RH conditions in adsorption and desorption. The results show hysteresis in both latewood and earlywood, with MC values in a range between 4\% and 18\% within the same RH range, that is, between 10\% and 85\%.

A recent investigation aimed at documenting the shrinkage of wood from green to dried state. The measurements were performed at the University Gent Center for X-ray Tomography, similar to the one described in Masschaele et al. (2007). A sample of wood freshly taken from a live tree was subjected to several decreasing steps in RH. Using appropriately segmented 3D dataset, a cell wall volumetric indicator is obtained by calculating the amount of material from the binary images at each RH.

\begin{figure}[htb] \centering{ \includegraphics[width=8.cm]{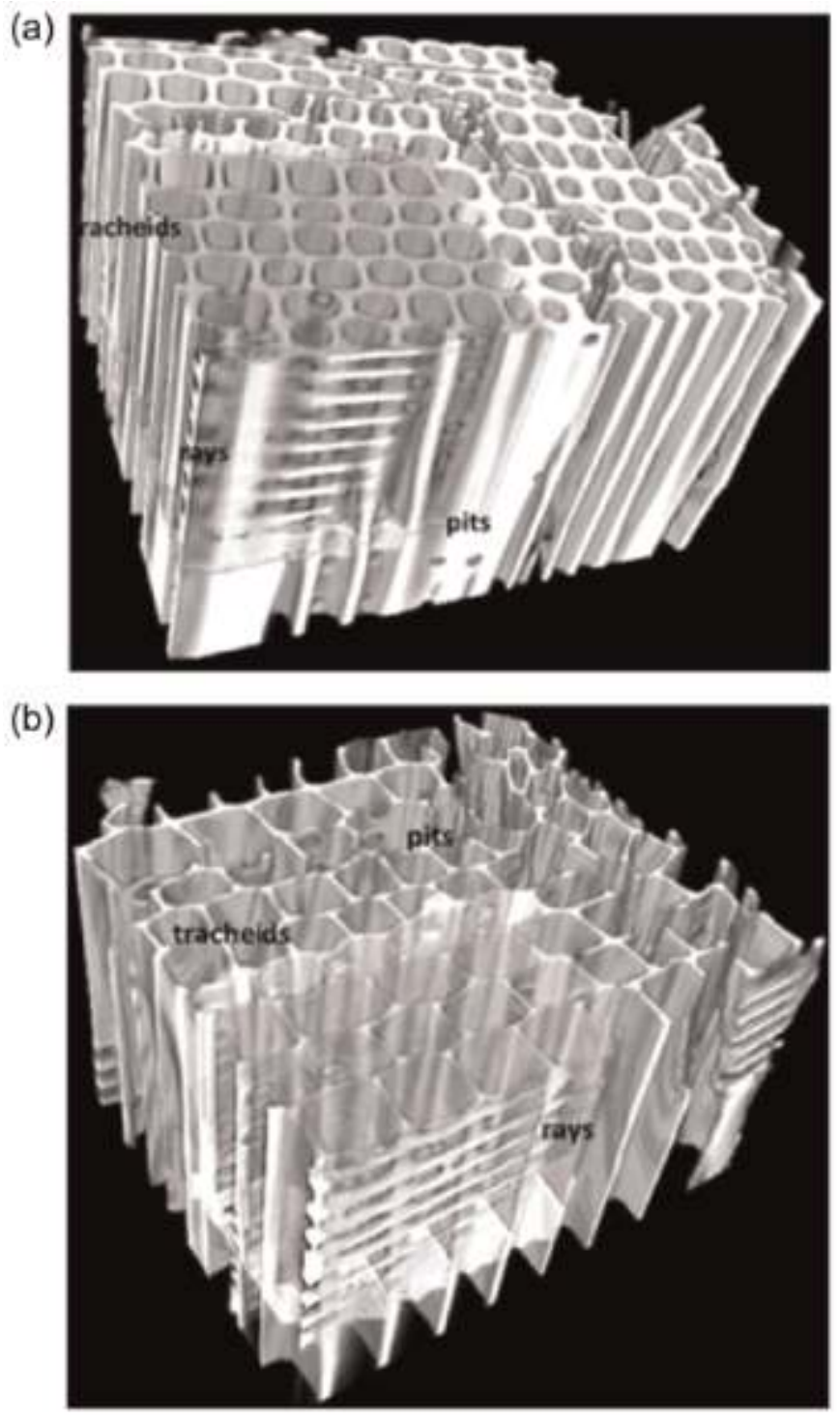} }
 \caption{Three-dimensional tomographic reconstruction of specimens: (a) latewood and (b) earlywood.} 
\end{figure}
In green state, a large percentage of the sample cell lumens are filled with liquid water. As wet wood dried, free water leaves the lumen and the swelling percentage is obtained by comparison of a volumetric indicator in relation to the smallest volume measures. Similar to the previous work, a DVS analysis was performed to study the MC from the green to the dried states. The behavior in the first desorption, and a subsequent sorption/desorption loop, is shown in Figure 13.

The series of tridimensional dataset during swelling and shrinkage is then used as geometrical input and for the validation of the multiscale modeling work presented above.

\begin{figure}[htb] \centering{ \includegraphics[width=12.cm]{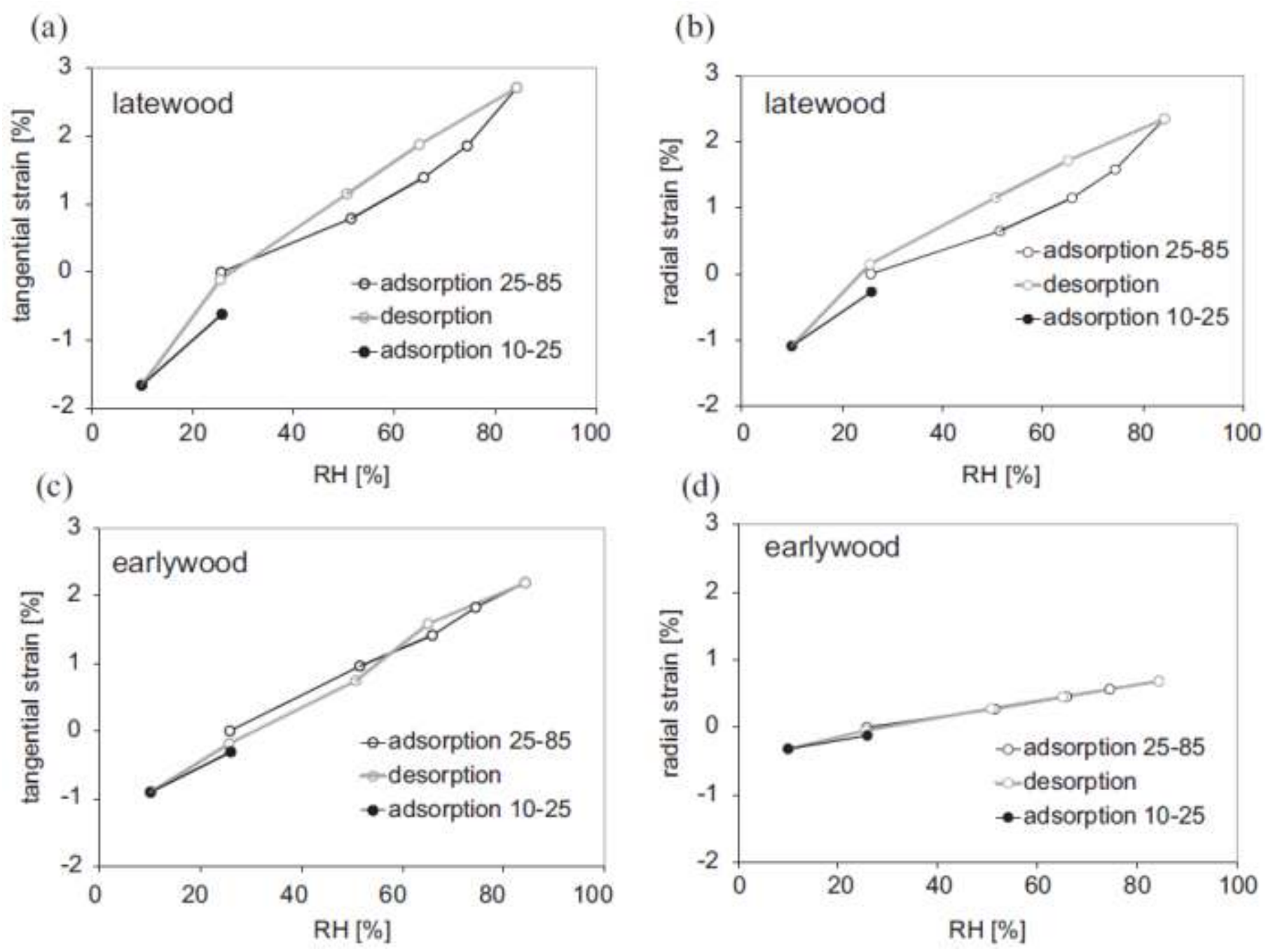} }
 \caption{Swelling strain measurements on pure latewood in (a) tangential and (b) radial directions and on pure earlywood in (c) tangential and (d) radial directions using microtomography dataset. RH: relative humidity.} 
\end{figure}
\begin{figure}[htb] \centering{ \includegraphics[width=8.cm]{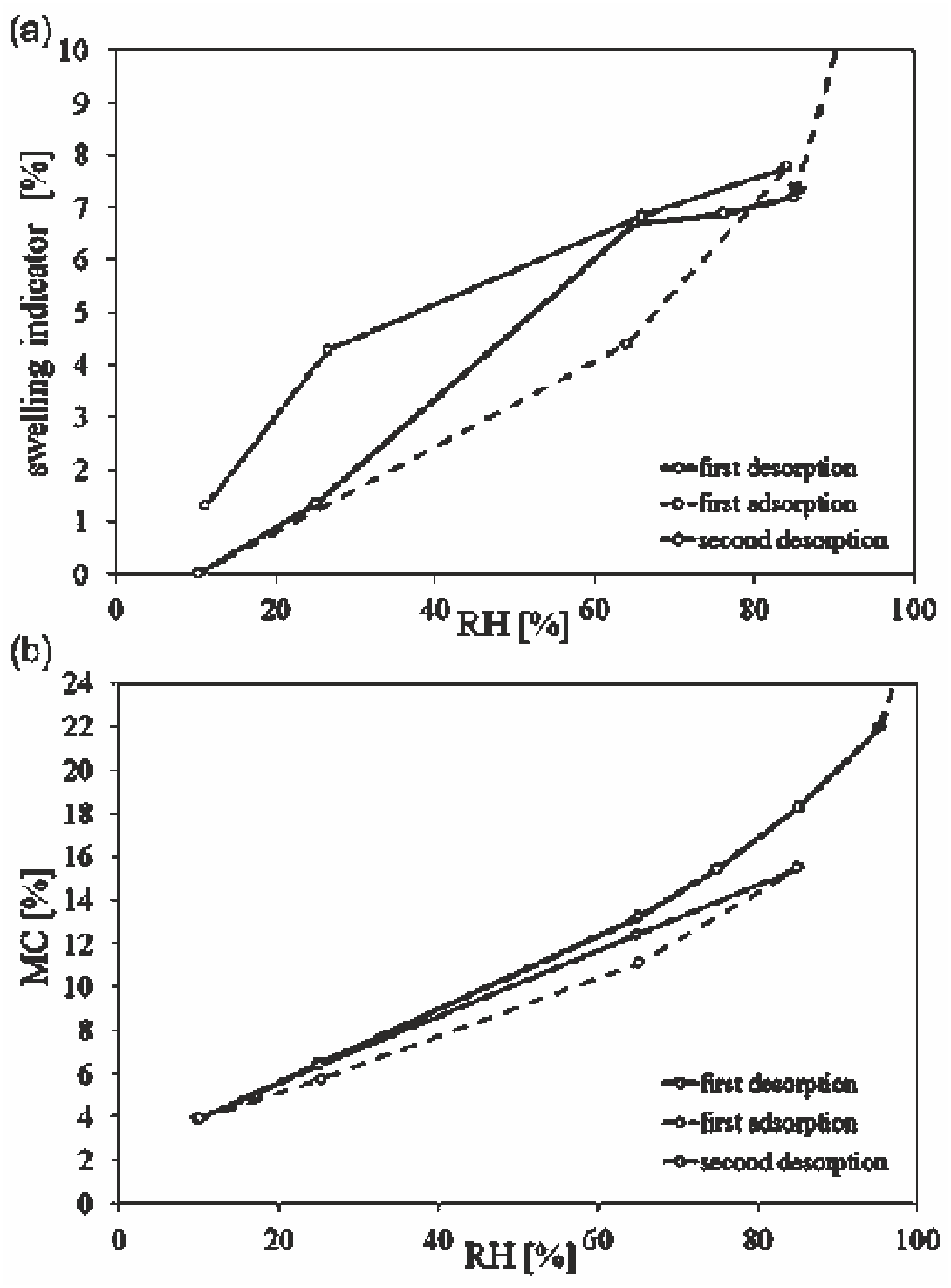} }
 \caption{(a) Swelling indicator versus RH for first desorption from green state and adsorption, and second desorption loop, for spruce. (b) MC versus RH for the same sample. RH: relative humidity; MC: moisture content.} 
\end{figure}
%%%
\subsection{Intraring variation of cross-sectional swelling by DIC}
\begin{figure}[htb] \centering{ \includegraphics[width=12.cm]{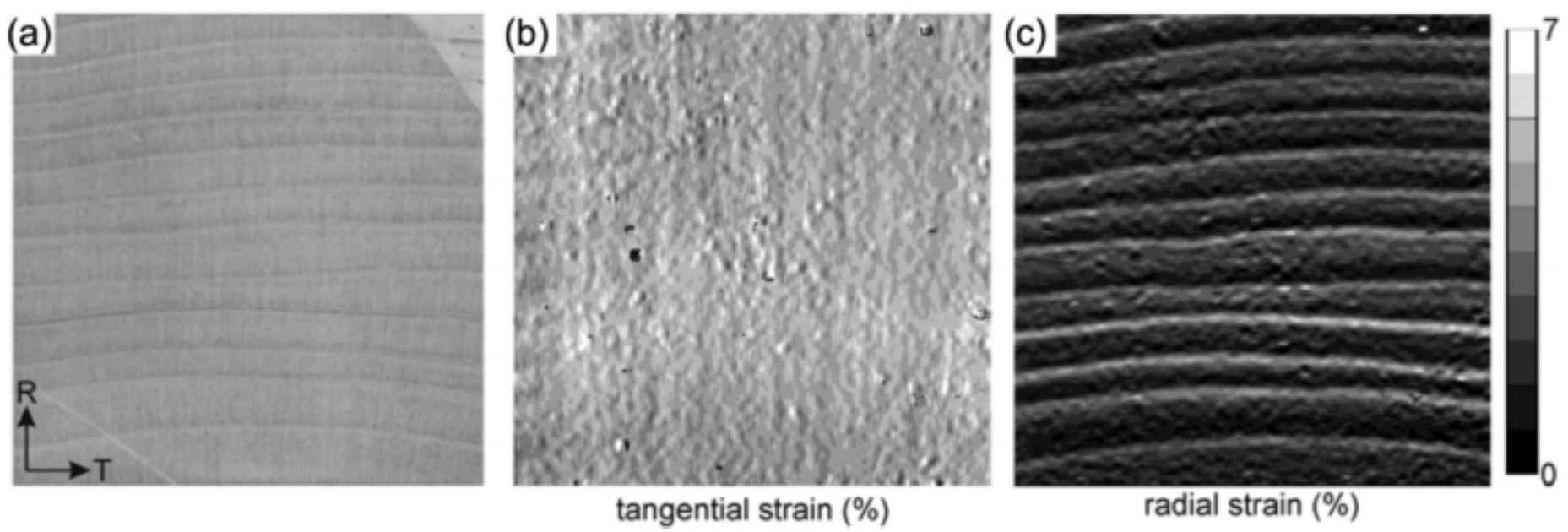} }
 \caption{Distribution of cross-sectional strains at moisture content 16.84\% of the sample seen in (a) orthographic image, (b) tangential direction, and (c) radial direction. Scale of legend is swelling strain (\%).} 
\end{figure}
In order to compare the dimensional changes as generated by X-ray tomography on segmented earlywood and latewood samples with swelling at the macroscale, we also investigated the moisture-induced deformation on the growth ring scale using a commercial DIC system (VIC2D; Correlated Solutions, USA). The noninvasive system enables to study the surface deformation of samples by cross-correlation of gray-value distributions. Therefore, a random gray-value speckle pattern was applied on the sample surfaces. The samples (40(R) x 40(T) x 5(L) mm$^3$) were then conditioned in a climatic chamber until they reached equilibrium, and an image of the sample was acquired. The images for each RH steps were then analyzed with the image correlation software and referenced to the dry-state image. 

The full deformation strain fields of a sample that underwent a MC change from oven dry (0\%) to 16.84\% are displayed in Figure 14. One can clearly see that the growth ring structure of the sample (Figure 14(a)) is clearly reflected in the radial strain distribution (Figure 14(c)), whereas the tangential strain field appears to be more or less homogeneous. In the radial strain distribution, the regions of higher deformation coincide with the latewood regions. Line extraction of radial and tangential strains is given in Figure 15. The average strain in radial direction is approximately 2.1\%, whereas the tangential average strain is 4.8\%, thus more than double. Furthermore, the more or less constant tangential swelling strain over the area of the sample results in a low deviation between maximum and minimum strains (2.5\%). For the radial strains, however, pronounced differences can be seen for earlywood and latewood resulting in the high deviation of 5.6\%. These marked differences coincide with the radial cell wall thickness and thus density profile along the growth ring where the thin-walled earlywood cells result in low density and the brick-shaped thick-walled latewood cells in a significantly higher density (e.g. Derome et al., 2011; Lanvermann et al., in press). Thus, we can see that at the macroscale, latewood undergoes similar strains in tangential and radial directions, while earlywood undergoes low radial swelling strains compared to what happens in the tangential direction, confirming the findings presented above on the isolated tissues documented with X-ray tomography. 
\begin{figure}[htb] \centering{ \includegraphics[width=8.cm]{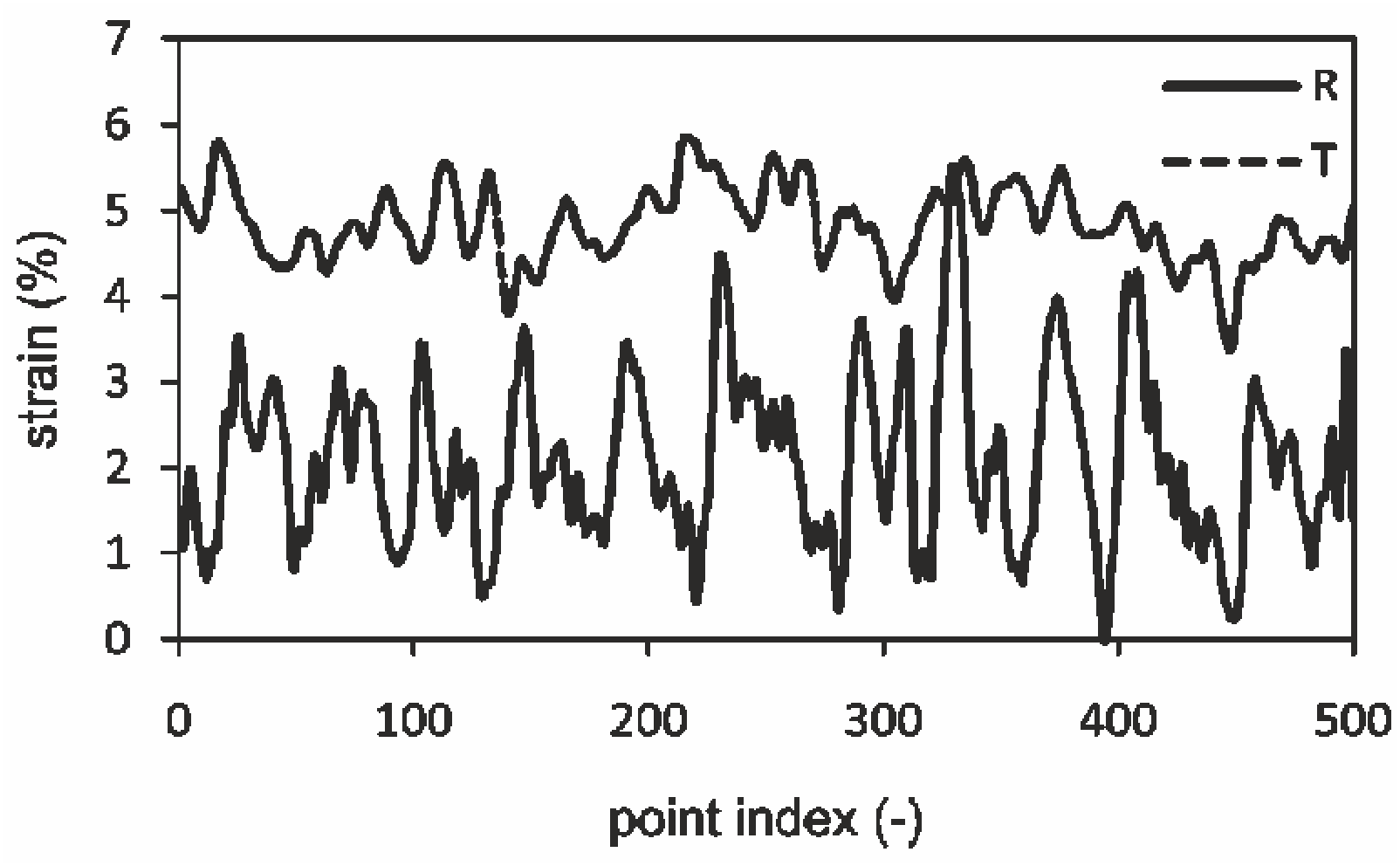} }
 \caption{Line extractions from the strain fields in radial and tangential directions. The radial strain distribution is displayed with solid line and the tangential one with a dashed line.} 
\end{figure}
%%%
\subsection{Neutron radiography for moisture transport analysis}
\begin{figure}[htb] \centering{ \includegraphics[width=8.cm]{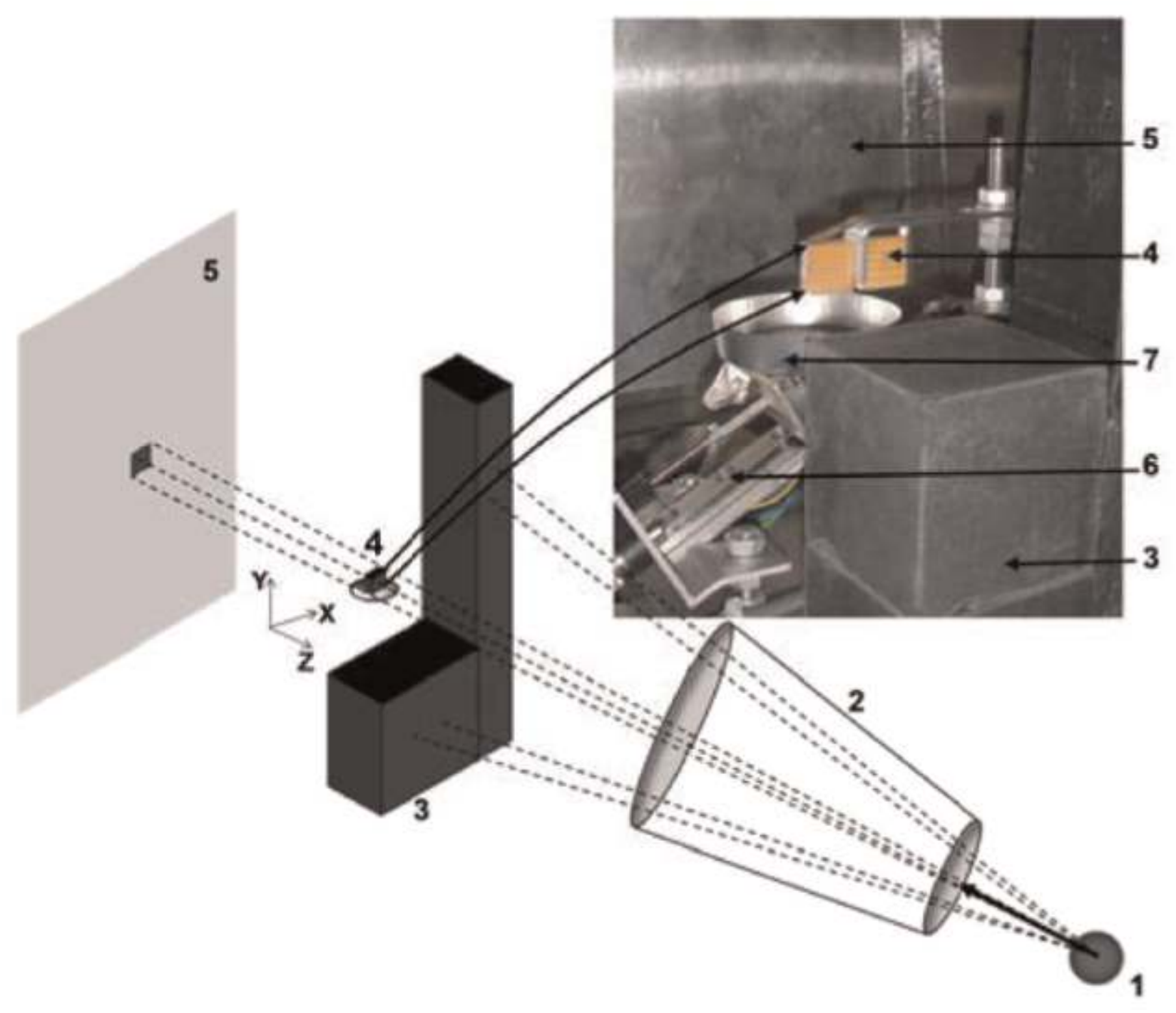} }
 \caption{Schematic overview of the experimental setup inside the NEUTRA beamline: (1) neutron source, (2) collimator, (3) boronated polyethelene blocks for shielding, (4) sample installed on the sample holder, (5) detector, (6) remote control elevator, and (7) water container.} 
\end{figure}
Liquid water uptake in three softwood species (spruce, fir, and pine) in the radial, tangential, and longitudinal directions, is investigated using neutron radiography at the wood growth ring scale. The high sensitivity of neutron to hydrogen atoms enables an accurate determination of the change in the MC of wood. The analysis of the spatial and temporal change of water content distribution shows that liquid water transport has different characteristics depending on the direction of uptake. Figure 16 shows our custom-made experimental setup inside the thermal neutron radiography facility. A reference radiograph of the specimens at the initial state (dry or moist) is obtained after the sample is placed in the sample holder. Then, the water container is remotely moved up until the base of the sample is in contact with the free water surface. From the time of contact with water, radiographies are acquired at 60-s intervals, over a period of 60-180 min depending on the uptake rate of the different specimens. 
\begin{figure}[htb] \centering{ \includegraphics[width=6.cm]{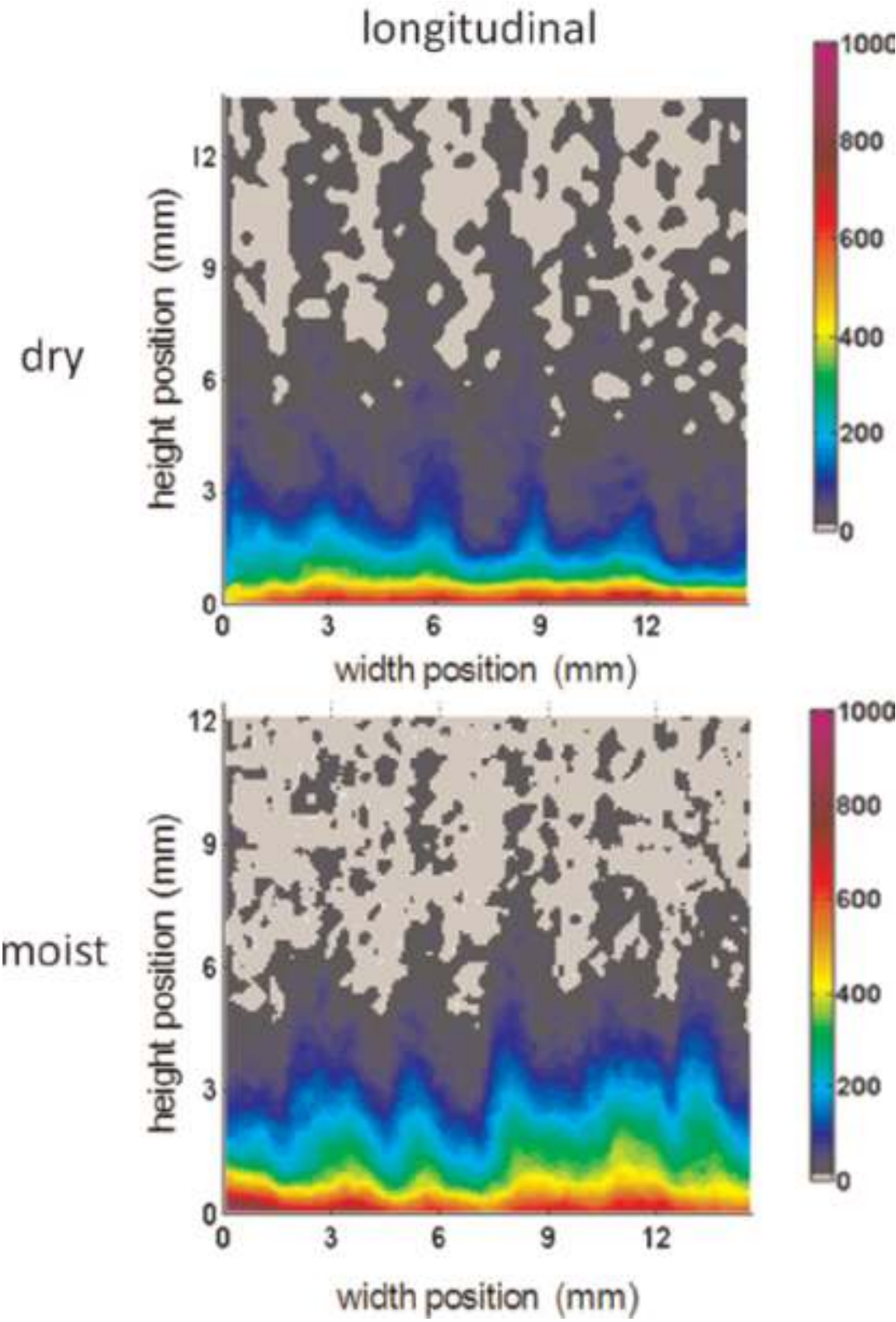} }
 \caption{Distribution of moisture content in fir specimens after 60 min of water uptake in longitudinal with dry or moist initial conditions (legend is in kg/m$^3$).} 
\end{figure}
\begin{figure}[htb] \centering{ \includegraphics[width=5.5cm]{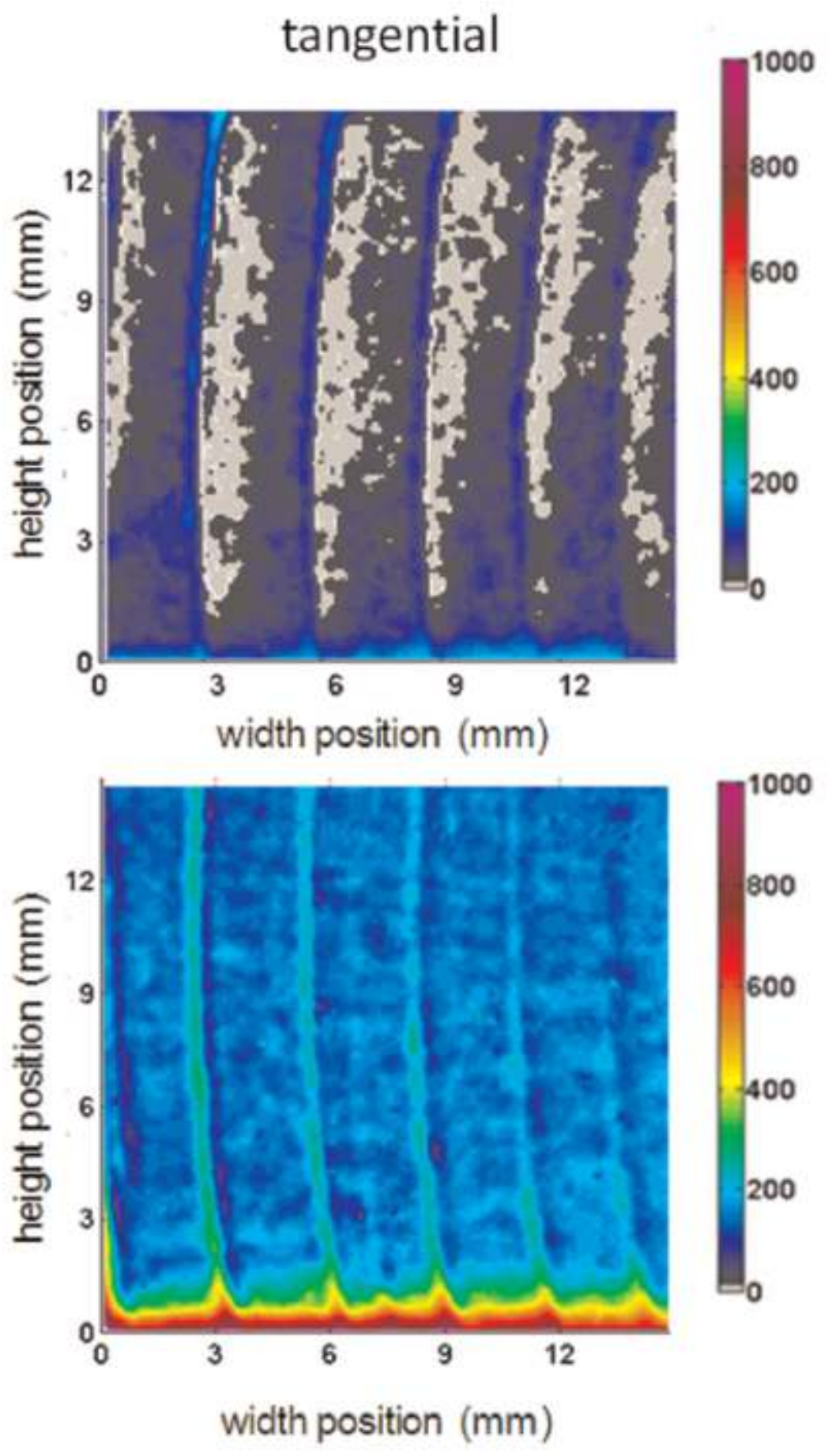} }
 \caption{Distribution of moisture content in fir specimens after 43 min of water uptake in tangential direction with dry or moist initial conditions.} 
\end{figure}
\begin{figure}[htb] \centering{ \includegraphics[width=6.cm]{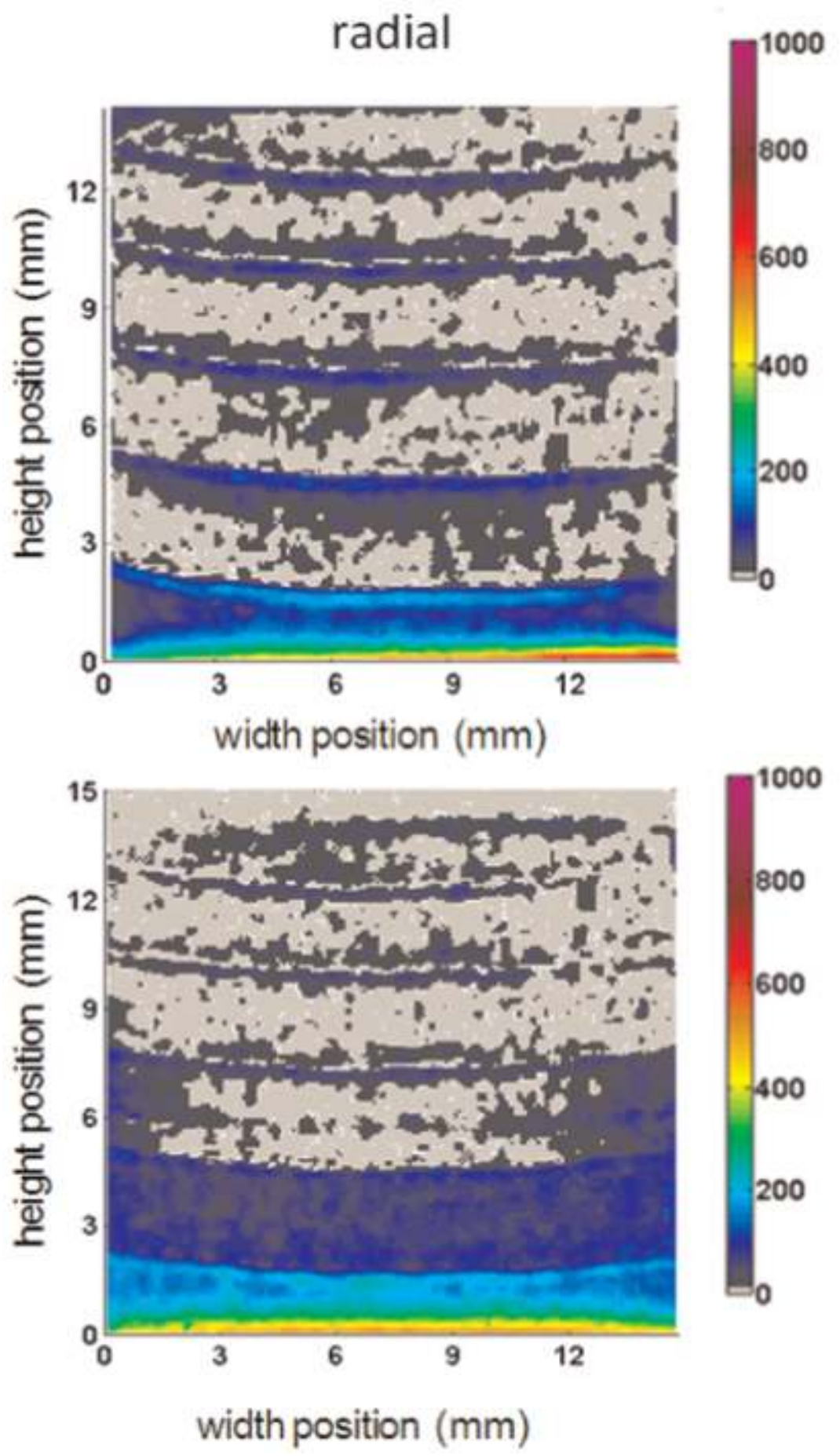} }
 \caption{Distribution of moisture content in radial direction after 145 min of water uptake with dry or moist initial conditions.} 
\end{figure}

As a control reference, the total mass of intruded water during the uptake test is measured by weighing the mass of the specimens during the uptake test. To validate the moisture uptake curves obtained from neutron radiographs, we also repeated a series of water uptake tests on the same samples in the laboratory using a precision balance. The detailed processing of the images is described in Sedighi Gilani et al. (2012). Our results, see examples in Figures 17 to 19, show that latewood cells play a more significant role in water uptake than earlywood cells and that ray tracheids also contribute to liquid transport. Latewood tracheids possess smaller cell lumens and a smaller ratio of aspirated bordered pits than earlywood cells that make them the preferential pathways for transport along the longitudinal direction. The process of liquid uptake is different in the radial and tangential directions as the path of the liquid is more intricate, involving also the rays and requiring more often traversing pits. In tangential direction, water uptake is occurring first in the latewood with a subsequent radial redistribution toward the earlywood. In radial direction, the growth ring boundary decreases the liquid transport rate, an indication that a significant portion of the rays are interrupted at that location. All liquid transport is accompanied by sorption in the cell walls, but this sorption is most visible during uptake in the radial direction. The moisture uptake rate in initially moist specimens is seen to be higher.

Neutron images can be processed to lead average MC along the height of the samples, shown in Figure 20(a), (c), and (e). Furthermore, neutron images can be registered to lead swelling coefficients as shown in Figure 20(b), (d), and (f).
\begin{figure}[htb] \centering{ \includegraphics[width=12.cm]{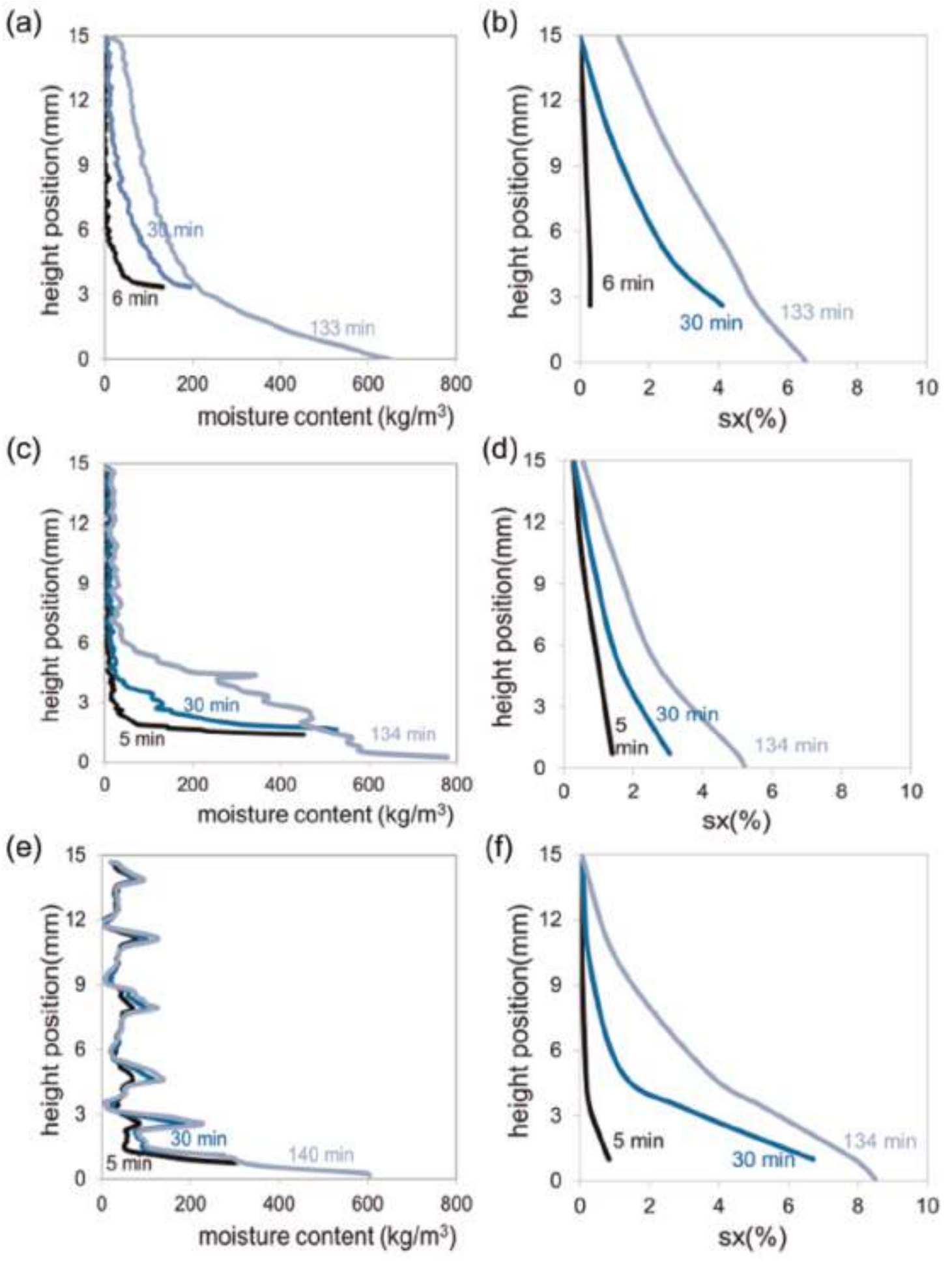} }
 \caption{Variation of moisture content along the height of pine samples during uptake in (a) longitudinal, (c) tangential, (e) radial directions, and (b, d, and f) corresponding swelling.} 
\end{figure}

The analysis of neutron images reveals interesting features about the process of liquid uptake and quantification and visualization of absorbed water, with high resolution (0.07 kg/m$^3$ in this experiment). We see a dependency of the liquid water transport on the hierarchical structure of softwood, orthotropic uptake directions, and initial MC of the specimens. Higher uptake of the initially moist specimens could be attributed to their already modified (swollen) geometry or the existence of thin film of water on the microscopic pore surfaces, which increase the rate of water absorption.
%%%%%%%%%%%%%%%%%
\section{Conclusion}
The proposed multiscale approach combines efficient computational tools with supporting swelling validation experiments at different spatial scales, as the mechanism of swelling links the moisture and mechanical behavior. The predicted swelling coefficients are compared to experimental data at the scales of the cell wall, of earlywood and latewood, and at macroscale. The agreement is good, predicting general trends accurately and giving a physical understanding of the observed phenomena. In addition, the complex structure of a natural material, such as wood, introduces a superposition of different influences that makes it difficult to evaluate. The modeling framework has the flexibility to include different nonlinear and moisture-dependent constitutive material behavior at the microlevel and can be used for studying the influence of morphological, mechanical, and moisture effects on effective hygromechanical properties, here, in softwoods and eventually in other wood-based composites. 

This comprehensive approach, combining modeling and advanced imaging, is expected to lead to a basic understanding of the cellular and hierarchic physics of wood in response to environmental stimuli, which could be a stepping stone for the understanding of other cellular or biological materials and the development of new biomimic materials with desired tailored properties.
%%%
\section*{Funding}
This study was supported by the Swiss National Science Foundation SNF under Sinergia grant no. 125184.
%%%
\section*{Acknowledgements}
Phase-contrast synchrotron X-ray tomographic data were acquired at the Tomcat beamline of SLS. Neutron images acquired at the Neutra beamline of SINQ. SLS and SINQ are at PSI, Villigen, Switzerland. X-ray tomography data for green wood experiments were acquired at the University Ghent Center for X-ray Tomography in Belgium.
%%%
\section*{References}
\begin{itemize}
\setlength{\itemsep}{12pt}
\item[] Biot MA (1941) General theory of three-dimensional consolidation. Journal of Applied Physics 12: 155-164.
\item[] Boutelje JB (1962) The relationship of structure to transverse anisotropy in wood with reference to shrinkage and elasticity. Holzforschung 16(2): 33-46.
\item[] Carmeliet J, Derome D, Dressler M, et al. (2013) Nonlinear poro-elastic model for unsaturated porous solids; Olivier Coussy Memorial Issu. Journal of Applied Mechanics 80(2): 020909.
\item[] Cave ID (1978) Modelling moisture-related mechanical properties of wood. Part I:properties of the wood constituents. Wood Science and Technology 12: 75-86.
\item[] Chamis CC (1984) Simplified composite micromechanics equations for hygral, thermal, and mechanical properties. Sampe Quarterly: Society for the Advancement of Material and Process Engineering 15(3): 14-23.
\item[] Coussy O (2010) Mechanics and Physics of Porous Solids. John Wiley and Sons, West Sussex, United Kingdom.
\item[] Derluyn H, Janssen H, Diepens J, et al. (2007) Hygroscopic behavior of paper and books. Journal of Building Physics 31(9): 9-34.
\item[] Derome D, Griffa M, Koebel M, et al. (2011) Hysteretic swelling of wood at cellular scale probed by phase-contrast X-ray tomography. Journal of Structural Biology 173: 180-190.
\item[] Derome D, Zillig W and Carmeliet J (2012) Variation of measured cross-sectional cell dimensions and calculated water vapor permeability across a single growth ring of spruce wood. Wood Science and Technology 46(5): 827-840.
\item[] Farruggia F and Perre P (2000) Microscopic tensile tests in the transverse plane of earlywood and latewood parts of spruce. Wood Science and Technology 34(2): 65-82.
\item[] Halpin JC and Kardos JL (1976) The Halpin-Tsai equations: a review. Polymer Engineering and Science 16(5): 344-352.
\item[] Lanvermann C, Evans R, Schmitt U, et al. (in press) Distribution of structure and lignin within growth rings of Norway spruce. Wood Science and Technology.
\item[] Masschaele BC, Cnudde V, Dierick M, et al. (2007) UGCT: new X-ray radiography and tomography facility. Nuclear Instruments and Methods in Physics Research Section A: Accelerators Spectrometers Detectors and Associated Equipment 580: 266-269.
\item[] Neuhaus FH (1981) Elastizitaetszahlen von Fichtenholz in Abaengigkeit von der Holzfeuchtigkeit. PhD Thesis, Institut fuer konstruktiven Ingenieurbau Ruhr-Universitaet Bochum, Germany.
\item[] Persson K (2000) Micromechanical Modelling of Wood and Fibre Properties. Doctoral Thesis, Lund University, Sweden.
\item[] Rafsanjani A, Derome D, Wittel F, et al. (2012) Computational up-scaling of anisotropic swelling and mechanical behavior of hierarchical cellular materials. Composites Science and Technology 72: 744-751.
\item[] Sedighi-Gilani M, Griffa M, Mannes D, et al. (2012) Visualization and quantification of liquid water transport in softwood by means of neutron radiography. International Journal of Heat and Mass Transfer 55 (21-22): 6211-6221.
\item[] Tsai S and Hahn H (1980) Introduction to Composite Materials. Lancaster, PA: Technomic Publishing Company, Inc.
\item[] Viitanen H, Vinha J, Salmiren K, et al. (2010) Moisture and bio-deterioration risk of building materials and structures. Journal of Building Physics 33(3): 201-224.
\item[] Watanabe U, Norimoto M and Morooka T (2000) Cell wall thickness and tangential Young's modulus in coniferous early wood. Journal of Wood Science 46(2): 109-114.
\end{itemize}
\end{document}